 \documentclass[12pt]{article}
 \usepackage{amsfonts,amssymb,amsmath}
  
  \newcommand{\Hide}[1]{}
  
 \newcommand{\End}{\nonumber\\ }
 \newcommand\Mbox[1]{ \mbox{\rm{#1}}}
 \newcommand\Mboxq[1]{\quad \mbox{\rm{#1}}\quad}
 
 \newcommand{\Textfrac}[2]{{\textstyle{\frac{#1}{#2}}}}
 
 \newcommand\Texthalf{\Textfrac{1}{2}}
 
 \newcommand{\At}[2]{\left.#1\right|_{#2}}
 
 \newcommand{\Lrb}[1]{\left( #1 \right)}
 
  \newcommand{\Set}[1]{\left\{ #1 \right\}}
  
 \newcommand{\Ipa}[2]{\left\langle #1, #2 \right\rangle}
  \newcommand{\Comm}[2]{\left[ #1 , #2 \right]}
  
 \newcommand{\Pb}[2]{\left\{ #1 , #2 \right\}}
 \newcommand{\Pbbl}[3]{\left\{ #1 , \left\{ #2,#3\right\} \right\}}
 \newcommand{\Pbbr}[3]{\left\{  \left\{#1 , #2\right\},#3 \right\}}
 
 \newcommand{\Exp}[1]{\exp\left(#1\right)}
 
 \newcommand\Real{ {\mathbb R} }
 
 \newcommand{\Df}{ {\mathrm{d}}}
 \newcommand{\Dbd}[1]{\frac{\Df}{\Df #1} }
 \newcommand{\Dbdt}[2]{\frac{\Df #1}{\Df #2} }
 
 \newcommand{\Pbd}[1]{\frac{\partial}{\partial #1} }
 \newcommand{\Pbdt}[2]{\frac{\partial #1}{\partial #2}}
 \newcommand{\Fbdt}[2]{\frac{\delta #1}{\delta #2}}
 
 \newcommand{\Dc}{\mathcal{D}}
 \newcommand{\Cbd}[1]{\frac{\Dc}{\Dc #1} }
 \newcommand{\Cbdt}[2]{\frac{\Dc #1}{\Dc #2}}
 
  \newcommand{\Al}{\alpha}
  
  \newcommand{\Ga}{\gamma}
  \newcommand{\De}{\delta}
 \newcommand{\Ep}{\epsilon}
 
 \newcommand{\Ka}{\kappa}
 \newcommand{\La}{\lambda}
 
 \newcommand{\Matt}[4]{\left(
        \begin{array}{cc}#1&#2\\#3&#4\end{array}\right)}
 
 \newcommand{\Lagr}{\mathfrak{L}}
 \newcommand{\Dldq}{\Pbdt{\Lagr(q(t),\Dotq(t))}{q^i(t)}}
 \newcommand{\Dldqdot}{\Pbdt{\Lagr(q(t),\Dotq(t))}{\Dotq^i(t)}}
 
 \newcommand{\Refeq}[1]{(\ref{#1})}

\newcommand{\Omit}[1]{}
 
 \newcommand{\Dotq}{\dot{q}}
 \newcommand{\Hame}{H_e}
 \newcommand{\Igen}{G(t,p_i(t),q^i(t))}
 \newcommand{\Gt}{g(p(t),q(t))}
 \newcommand{\Man}{\mathcal{M}}
 
 \newcommand{\Manp}{\tilde{\mathcal{M}}}
  \newcommand{\Mex}{\bar{\mathcal{M}}}
 \newcommand{\Intt}{\int_{0}^{T}}
 \newcommand{\Cinf}{C^{\infty}{}}
 \newcommand{\Omp}{\tilde{\omega}}
 \newcommand{\Groupp}{\tilde{G}}
 \newcommand{\Jp}{\tilde{J}}
 \newcommand{\Lieg}{\mathfrak{g}}
\newcommand{\Liegd}{\mathfrak{g}^*}
 \newcommand{\Liegp}{\tilde{\mathfrak{g}}}
 \newcommand{\Liegdp}{\tilde{\mathfrak{g}}^*}
 \newcommand{\Realp}{\tilde{\Real}}
 \newcommand{\Darg}[1]{#1(\cdot)}
 \newcommand{\Const}{\phi}
 \newcommand{\Prim}{\phi^{(1)}}
 \newcommand{\Seco}{\phi^{(2)}}
 \newcommand{\Cons}{\psi}
 
 \newcommand{\Hamflow}[2]{\rho^H_{#1}(#2)}
 
 \newcommand{\Hamden}{\mathcal{H}} 
 \newcommand{\Eps}{\epsilon}

\newcommand{\ud}{\mathrm{d}}
\newcommand{\um}{\mathrm{m}}
\newcommand{\xd}{\dot{x}_{\mu}}
\newcommand{\xu}{\dot{x}^{\mu}}

\newcommand{\fpartial}[2]{\frac{\partial #1 }{\partial #2 }}
\newcommand{\Xd}[1]{\dot{X}^{#1}}
\newcommand{\Xp}[1]{X'^{#1}}
\newcommand{\gumn}{g^{\mu\nu}}
\newcommand{\gdmn}{g_{\mu\nu}}
\newcommand{\prim}[1]{\phi^{\left( 1\right) }_{#1}}
\newcommand{\secu}[1]{\phi^{\left( 2\right) }_{#1}}
\begin{document}

 \begin{center}
 {\large BRST cohomology of systems with secondary constraints}\\
 \ \\
   Alicia Lopez and Alice Rogers\\
  \ \\
  Department of Mathematics      \\
  King's College             \\
  Strand, London  WC2R 2LS         \\
 \end{center}

 \vskip0.2in
 \begin{center}
 \today
 \end{center}
 \vskip0.2in
 \begin{abstract}
Extending  phase space to include time and it canonical conjugate energy as well as the usual momentum and position variables, and then introducing the constraint which sets energy equal to the Hamiltonian, gives a symplectic action of the additive group $\Real$ which corresponds to time translation along the solutions to the equations of motion. This allows the BRST operator of a system with secondary constraints to be constructed from first principles.
 \end{abstract}
\section{Introduction}

This paper derives from first principles the extra ghost terms which must be added to the classical Hamiltonian in the BRST quantization of systems with secondary (and higher) constraints.

BRST cohomology  provides a systematic approach to the path integral quantization of systems with symmetry, effectively reducing the integrals to the true phase space of the system even when this is a finite-dimensional moduli space.  In the Hamiltonian setting the symmetries of a system manifest themselves as constraints on the phase space, as was first shown by Dirac \cite{Dirac}. For systems with first class constraints all of which are primary, the BRST method is well understood and can be justified in terms of Hamiltonian group actions on symplectic manifolds \cite{FigKim1991,HenTei,KosSte,GFBFVQ}.
However for many realistic systems the situation is more complicated, in particular when a system has secondary constraints not only must the correct BRST operator be constructed, but also an extension of the Hamiltonian is required so that it commutes with the BRST operator.  This paper extends the understanding of BRST quantization to systems with secondary (and higher) constraints, deriving the extra terms in the Hamiltonian from first principles (as opposed to simply adding terms so that the Hamiltonian commutes with the BRST operator) by extending phase space to include time and its canonical conjugate energy.

In the Lagrangian setting it was recognised by Noether that a conserved current could be associated with each symmetry generator of a system. A description of these ideas and their historical development  may be found in Kosmann-Shwarzbach \cite{Kosman}.  In his pioneering work  on the canonical quantization of systems with symmetries \cite{Dirac} Dirac  recognised that in the canonical setting symmetries manifest themselves as constraints on phase space. In the simplest cases a Lagrangian system with $n$ degrees of freedom whose symmetries give rise to $m$ Noether currents leads to a phase space with $m$ primary constraints which are both first class and conserved in time; these primary constraints correspond directly to the Noether currents, and lead via the Marsden-Weinstein reduction process \cite{MarWei} to the true phase space of the system which has dimension $2n-2m$.  However it was observed by Dirac
that systems occur whose primary constraints are not constants of the motion and thus do not directly correspond to the Noether currents of the theory. The time variation of the primary constraints are normally introduced as secondary constraints, but at the expense of the standard machinery which has been developed using constraints to understand the reduced or true phase space of the theory and hence proceed to quantization.

This paper provides an understanding of such systems, by considering the space of phase space
trajectories rather than simply phase space, building on the work of Castellani, Henneaux, Tetelboim, Zanelli,  and Pons \cite{Castel82,HenTeiZan90,Pons05}.   A key point is that it is not appropriate to
regard primary and secondary constraints as individual constraints on phase space, instead
each primary constraint together with any corresponding secondary constraint must be built
into a single object on the space of trajectories in phase space; these composite objects
correspond to the momentum map of a Hamiltonian group action on the space of phase space
trajectories. In more mathematical language, the theory is developed in terms of the space of
paths in a symplectic manifold; this space itself has a symplectic structure, and it will be
shown that more general gauge symmetries, of the kind which lead to secondary (and possibly
tertiary and so on) constraints can be formulated by extending standard constructions in
Hamiltonian dynamics to the symplectic manifold of paths.

A further important step is the extension of phase space to include time and its canonical conjugate energy, as described in Woodhouse \cite{Woodho}. Imposing the constraint that the energy is equal to the Hamiltonian function then allows us to take a `snapshot' in the space of classical trajectories in a manner which respects the dynamics of the system, so that a finite-dimensional phase space can be used rather than the infinite-dimensional space of trajectories, as is explained in section \ref{secQUANTBRST}.

In section \ref{secSTANDARD} the standard BRST construction for systems with primary first class constraints is briefly reviewed. Important steps towards understanding the structures which arise when secondary constraints
occur made by a number of authors including Castellani, Henneaux, Tetelboim, Zanelli,  and Pons \cite{Castel82,HenTeiZan90,Pons05} are is described section \ref{secPONS} with particular emphasis on the work of Pons who recognised that Dirac had restricted his analysis to
an infinitesimal neighbourhood of the initial conditions, while it was in fact necessary to
consider the gauge parameters and phase space trajectories for all values of the time
parameter $t$.

In section \ref{secPaths}, using rather more mathematical language and largely following \cite{Wurzba95}, we
review the symplectic geometry of the space of paths in phase space,
and show how a Hamiltonian group action on this space leads to a
momentum maps that encodes primary and secondary constraints in a
single object and allows a finite-dimensional formalism to be used. In section  \ref{secQUANTBRST} the BRST operator, as well as the extended Hamiltonian,  is constructed for systems with secondary constraints, and the full quantization procedure described. This is achieved by extending phase space to include time $t$  and energy $E$, together with the additional constraint $H-E=0$ which corresponds to time translation along the solutions to the equations of motion.

To set the scene we will describe a toy example introduced in \cite{DreFisGreHen} which gives a simple
illustration of the issues involved and establishes notation. (Several more examples are analysed in the final section.) Consider
a system with two degrees of freedom $q^i,i=1,\dots,2$ and Lagrangian
 \begin{equation}\label{eqTOYLAG}
    \Lagr \Lrb{q^i,\Dotq^i} =  \Texthalf(\Dotq^2+q^1)^2
       \, .
 \end{equation}
The action is invariant under transformations
 \begin{equation}\label{eqTOYINV}
    \De(q^2)=\Ep \Mboxq{and} \De(q^1)=-\dot{\Ep}
 \end{equation}
for arbitrary $\Ep(t)$ and the equations of motion are
 \begin{equation}\label{eqTOYEOM}
        \dot{q}^2+{q}^1=0 \Mboxq{and} \ddot{q}^2+\dot{q}^1=0 \,.
 \end{equation}
 The Noether current corresponding to the invariance (\ref{eqTOYINV}) is
 \begin{equation}
    J(q^i,\Dotq^i) = \Dotq^2+q^1 \, ,
 \end{equation}
which is readily seen to be conserved provided that the equations of motion hold.

 Passing, via
the Legendre transformation, to the Hamiltonian formulation of this model we obtain
 \begin{equation}
    p_1 = \Pbdt{\Lagr}{\Dotq^1}= 0\, , \quad
     p_2 = \Pbdt{\Lagr}{\Dotq^2}=\Dotq^2 +q^1\,,
 \end{equation}
 so that we have one primary constraint
  $\Prim \equiv p_1 = 0$.  The
 Hamiltonian of the system is
  \begin{equation}
    H(p_i,q^i) =  + \Texthalf p_2{}^2  - p_2 q^1
  \end{equation}
 from which we see that the equation of motion for the constraint
  $\Prim \equiv p_1$ is
  \begin{equation}
      \dot{p}_1 =p_2 \, ,
  \end{equation}
so that this constraint  is not preserved in
time unless the secondary constraint
 $\Seco \equiv \dot{p_1}=p_2=0$ also holds.  This secondary constraint is preserved in time so that
there is no tertiary or higher constraint.  One purpose of this paper
will be to reconcile the apparent mismatch between the number of Noether currents (in this case one) and constraints (in this case two) by
building primary and secondary (and where they exist tertiary and so on) constraints into a single object.

Throughout this paper we assume that there are no second class constraints or reducible symmetries.  To include them would introduce further complication without throwing more light on the developments in this paper. We also use the language of quantum mechanics rather than field theory, although field theoretic examples are considered in section \ref{secEXAMPLES}.

\section{Review of case where there are no secondary constraints}
\label{secSTANDARD}
For systems which simply have primary first class constraints there is a full
understanding of the role of symmetry in the classical dynamics which allows path integral quantization of the system using BRST techniques. These ideas are reviewed in this section, using the formalism of quantum mechanics.

Suppose that an $n$-dimensional system with position variables $q^i(t), i=1, \dots,n$ has
action
 \begin{equation}
    S\Lrb{q(\cdot),\Dotq(\cdot)} = \int_{0}^{T}\Lagr \Lrb{q(t),\Dotq(t)} \Df t
 \end{equation}
 and that this action is invariant if $q^i$ changes by
  \begin{equation}
    \delta q^i (t)= \Ep^a(t) R_a^i (q(t))
  \end{equation}
where for $a=1, \dots , m$ each $\Ep^a$ is an arbitrary and independent function of time $t$ and the summation convention has been used,  so that
\begin{eqnarray}\label{eqDELTAR}
  0 &=& \delta S \End
   &=& \int_0^T \Ep^a(t) R^i_a (q(t)) \Dldq \End
   && + \Lrb{\dot{\Ep}^a(t) R_a^i(q(t)) + \Ep^a(t) \Dbd{t}{R^i_a (q(t))}} \Dldqdot \, \Df t \,.
\end{eqnarray}
 Then, since each $\Ep^a$ is an arbitrary function of $t$,
\begin{eqnarray}
  R^i_a(q(t)) \Dldqdot  &=& 0\label{eqCON}\End
  &&\\
  \mbox{and}\quad\,\, R^i_a(q(t)) \Dldq
   + \Dbd{t}R^i_a(q(t)) \Dldqdot
   &=& 0 \,.\End\label{eqCUR}
\end{eqnarray}
From (\ref{eqCON}) we deduce that on taking the Legendre transformation
there will be constraints
 \begin{equation}
    \Const_a (p,q)\equiv R_a^i(q) p_i  = 0 \, \qquad a=1,\dots,m
 \end{equation}
while (\ref{eqCUR}) shows that, when the position variables $q^i(t)$ satisfy the equations of
motion
 \begin{equation}\label{eqEOMLAG}
    \Dbd{t} \Dldqdot = \Dldq \, ,
 \end{equation}
the Noether current $J_a \equiv R^i_a(q(t)) \displaystyle{\Dldqdot} $ is
conserved, so that, if $\Lrb{p_i(t),q^i(t)}$ is a trajectory in phase space which satisfies the equations of motion, then
 \begin{equation}\label{eqNOSEC}
    \Dbd{t}\Lrb{\Const_a(p_i(t),q^i(t))} = 0 \, .
 \end{equation}

   Standard arguments show
 that the equations of motion (\ref{eqEOMLAG}) for the Lagrangian system
  may be regained from the equations of motion of the system
 with extended Hamiltonian
  \begin{equation}\label{eqHAME}
    \Hame(p,q,\La) = H(p,q)
    +  \La^a \Const_a(p_i,q^i)
  \end{equation}
where the variables $\La^a, a= 1,\dots, m $ are Lagrange multipliers
for the constraints and $H$ is the canonical Hamiltonian
 \begin{equation}
     H(p,q) = \dot{q}^i p_i -  \Lagr \Lrb{q,\Dotq}\,.
 \end{equation}

In the Hamiltonian setting (\ref{eqNOSEC}) corresponds to
$\Pb{H}{\Const_a}=0$\,. If additionally the constraints satisfy a relation of the form
 \begin{equation}\label{eqCONALG}
    \Pb{\Const_a}{\Const_b} = C_{ab}^c \Const_c
 \end{equation}
 then they correspond to the momentum map of an infinitesimal Hamiltonian action on the phase space
 $\Real^{2n} \cong T^*(\Real^n)$
 of a Lie group $G$
 whose Lie algebra referred to a particular basis takes the form (\ref{eqCONALG}). The constraints $\Const_a$ generate
 infinitesimal transformations on phase space which map a solution of the equations of motion of the system
  to another solution. Such transformations are known as gauge transformations and make explicit the redundancy of the
  system which follows from the symmetries in the Lagrangian.
 Assuming that there is actually a Hamiltonian action of the full group $G$ on the phase space, then the reduced phase space,
 which carries the true degrees of freedom of the system, is $C/G$ where $C$ is the submanifold of
 $\Real^{2n}$ where $\Const_a(p,q)=0, a=1, \dots, m$. A standard result of Marsden and Weinstein  \cite{MarWei}
 shows that $C/G$ has
 a natural symplectic structure. (The corresponding
  Poisson brackets correspond to Dirac brackets if gauge-fixing functions exist.)
  In this ideal situation the BRST procedure together with appropriate gauge-fixing leads to
a path integral quantization of the system in a manner which is well understood
\cite{KosSte,GFBFVQ,BOOK}. The BRST charge is
 \begin{equation}
    \Omega = \eta^a \Const_a + \Textfrac 12 \eta^a \eta^b \pi_c C_{ab}^{c}
 \end{equation}
where the ghosts $\eta^a$ and ghost momenta $\pi_a$ are canonically conjugate anticommuting coordinate functions on a super extension of the original phase space with super symplectic form   typically 
 \begin{equation}
    \Df p_i \wedge \Df q^i + \Df \pi_a \wedge \Df \eta^a \, .
 \end{equation}
  (This construction can of course
be expressed in intrinsic language.)  The space of observables on the true phase space $C/G$
can then be identified with the cohomology of $\Omega$ at ghost number zero (where each
ghost $\eta^a$ has ghost number $1$ and each ghost momentum $\pi_a$ has ghost number $-1$).  The quantization of the theory can be carried
out using path integrals in the super phase space using the extended Hamiltonian
 $H + \Comm{\Omega}{\chi}$ where $\chi$ is a gauge-fixing fermion for the theory.
Path integration leads to a supertrace which can be shown to give a trace over BRST cohomology provided that $\chi$ is chosen in such a way that the cohomologically trivial term $\Comm{\Omega}{\chi}$\,regulates the infinite series in these traces \cite{GFBFVQ}.

 While the procedure described above gives the basic principle of the BRST construction and its use to generate path integrals, many systems require an extension of these ideas.  In the following sections the construction is extended to systems with secondary constraints, where it is necessary to consider the space of trajectories in phase space.

 \section{Gauge generators and secondary constraints}\label{secPONS}
In this section systems whose symmetry leads to secondary constraints are considered.
Starting from the Lagrangian picture, where the nature of symmetries which lead to secondary constraints is identified,
the Noether currents and primary constraints (which in this case are not the same) are identified, and then,
largely following Pons \cite{Pons05}, the canonical analysis of such systems is described, and a key condition (\ref{eqPONS})
determining symmetries of the equations of motion is derived. Finally it is shown how solutions of this condition may be derived.  This section draws on work by Castellani \cite{Castel82} and by Henneaux, Teitelboim and Zanelli \cite{HenTeiZan90}, as well as the work of Pons already cited.

 Suppose that as before we have a classical system with $n$ degrees of freedom
 $q^i, i=1, \dots , n$ and Lagrangian $\Lagr(q,\Dotq)$, but that the action
  \begin{equation}
    S\Lrb{q(\cdot),\Dotq(\cdot)} = \int_{0}^{T}\Lagr \Lrb{q(t),\Dotq(t)} \Df t
 \end{equation}
is invariant when the infinitesimal change in the variables $q^i$ takes the form
 \begin{equation}
     \delta q^i (t)= 
     \Lrb{\Ep^a(t) R_a^i (q(t)) + \dot{\Ep}^a(t) S_a^i (q(t))} \, ,
 \end{equation}
so that both  arbitrary infinitesimal parameters $\Ep^a$ and their time derivatives
are involved.  In this situation the equivalent of (\ref{eqDELTAR}) is
 \begin{eqnarray}
  0 &=& \delta S \End
   &=& \int_0^T \Lrb{\Ep^a(t) R^i_a (q(t)) + \dot{\Ep}^a S^i_a (q(t))}\Dldq \End
   && \quad + \quad\Bigl(\dot{\Ep}^a(t) R_a^i(q(t)) + \Ep^a(t) \Dbd{t}{R^i_a (q(t))}\End
    &&\qquad+\quad\ddot{\Ep}^a(t) S_a^i(q(t)) + \dot{\Ep}^a(t) \Dbd{t}{S^i_a (q(t))}\Bigr)\Dldqdot \, .
\end{eqnarray}

 Taking coefficients of $\ddot{\Ep}$, $\dot{\Ep}$ and $\Ep$ in turn we deduce that
   \begin{eqnarray}
   S_a^i(q(t)) \Dldqdot &=& 0 \label{eqDELTARS1}\\
   S_a^i(q(t)) \Dldq + \Lrb{R_a^i(q(t)) + \Dbd{t} S^i_a(q(t))}\Dldqdot &=& 0 \label{eqDELTARS2}\\
   \Mbox{and} \quad R^i_a(q(t)) \Dldq + \Dbd{t} R^i_a(q(t)) \Dldqdot &=& 0 \, .\End
   \label{eqDELTARS3}
 \end{eqnarray}
 From (\ref{eqDELTARS3}) we deduce that when the equations of motion (\ref{eqEOMLAG}) hold
 \begin{equation}\label{eqNOETHER}
   \Dbd{t}\Lrb{ R^i_a(q(t)) \Dldqdot } = 0
 \end{equation}
 so that the Noether current  $R^i_a(q(t)) \Dldqdot$  is conserved on shell as before,
 while (\ref{eqDELTARS1}) shows that on passing to phase space using the Legendre transformation there are primary constraints
 \begin{equation}\label{eqPRIMARY}
    \Prim_a \equiv S^i_a(q)p_i=0
 \end{equation}
 but from (\ref{eqDELTARS2}) we see that these are not preserved in time by the equations of motion,
 instead we have
 \begin{equation}\label{eqSECONDARY}
    \Dbd{t}\Lrb{\Prim_a(p(t),q(t))} = \Seco_a(p(t),q(t))
 \end{equation}
 where $\Seco_a(p,q) \equiv R^i_a(q) p_i$.  The system thus has secondary constraints $\Seco_a(p,q)$ which,
 by (\ref{eqNOETHER}), are
 preserved in time by the equations of motion, so that there are no tertiary or higher constraints.
This analysis makes it clear that the signature of a system which
leads to secondary constraints in the Hamiltonian setting is that the infinitesimal variations in the paths $q^i(t)$
which are symmetries of the action involve time derivatives of the infinitesimal parameter. (Extending the arguments above shows that if the  action is invariant under changes in the fields which involve second derivatives of the parameters with respect to time, then there will be tertiary constraints, and so on.  For simplicity in the remainder of this paper we will concentrate on systems with primary and secondary constraints but no higher order constraints.)

  As before the equations of motion for the Lagrangian system (\ref{eqEOMLAG})
  may be regained from the equations of motion of the system
 with Hamiltonian
  \begin{equation}
    \Hame(p,q,\La) =  p_i \Dotq^i
    - \Lagr(q^i,\Dotq^i)
    +  \La^a \Prim_a(p,q) \, .
  \end{equation}
  (It is important to note that this
extended Hamiltonian does not include the secondary constraints of the
system, a point made very clearly by Pons \cite{Pons05}.) The equation of motion for $\La^a(t)$ is of course
 \begin{equation}
    \Prim_a(p(t),q(t))= 0
 \end{equation}
so that the condition that the primary constraints be zero is included in the equations of motion. From
(\ref{eqPRIMARY}) and (\ref{eqSECONDARY}) we deduce that
 \begin{equation}\label{eqFIRSTCLASS}
    \Pb{\Hame}{\Prim_a} = \Seco_a \Mboxq{and} \Pb{\Hame}{\Seco_a}=0\,.
 \end{equation}

In the case of the toy model \Refeq{eqTOYLAG} the extended Hamiltonian
takes the form
 \begin{equation}
    \Hame(p_1,p_2,q^1,q^2,\La)
    = \Texthalf (p_2)^2 -p_2q^1+\La p_1
 \end{equation}
 leading to equations of motion
 \begin{eqnarray}
 \label{eqTOYEOMHAM}
      \dot{p}_2=0, \qquad \Dotq^2 &=& p_2-q^1\,, \End
   \dot{p_1}= p_2, \qquad \Dotq^1 &=& \La\,, \End
   p_1 &=& 0
 \end{eqnarray}
which are equivalent to \Refeq{eqTOYEOM}.

  In many cases a covariant derivative is more appropriate, with the infinitesimal symmetry transformation
 of the path $q^i(t)$ expressed as
 \begin{equation}
    \delta q^i (t)=
    \Ep^a(t) R_a^i (q(t)) + \Cbdt{\Ep^a(t)}{t} S_a^i (q(t)) \,
 \end{equation}
 where
 $$\Cbdt{\Ep^a(t)}{t}= \Dbdt{\Ep^a(t)}{t} - M^a_b(q(t))\Ep^b$$
 for some connection $M$.  In this situation analogous  arguments to those leading to (\ref{eqPRIMARY}), (\ref{eqNOETHER})
 and (\ref{eqSECONDARY}) give
 \begin{equation}\label{eqNOETHERC}
   \Cbd{t}\Lrb{ R^i_a(q(t)) \Dldq } = 0
 \end{equation}
 so that the Noether current  $R^i_a(q(t)) \Dldq$  is covariantly conserved on shell,
 while as before we have the primary constraints $\Prim_a\equiv S^i_a(q(t))p_i=0$ which satisfy
 \begin{equation}\label{eqSECONDARYC}
    \Dbd{t}\Lrb{\Prim_a(p(t),q(t))} = \Seco_a(p(t),q(t))
 \end{equation}
 where $\Seco_a\equiv R^i_a(q(t))p_i$.
Additionally equations (\ref{eqFIRSTCLASS}) now have the form
\begin{equation}\label{eqFIRSTCLASSC}
    \Pb{\Hame}{\Prim_a} = \Seco_a   \Mboxq{and} \Pb{\Hame}{\Seco_a}= -M_a^b\, \Seco_b\,.
 \end{equation}

We will now investigate the gauge transformations which this system admits. Gauge
transformations are mappings of the space of trajectories in phase space into itself which map
each solution of the equations of motion with given initial conditions
 $p^i(0)=p_{i \,0}, q^i(0)=q^i_0$ for the
variables
$p_i(t),q^i(t)$ to another satisfying the same initial conditions. Such solutions are not unique
because of the presence of the Lagrange multipliers in the equations of motion.  (In the next section it is shown that such transformations are canonical in the symplectic manifold of paths in phase space,
 with infinitesimal generators acting
by Poisson bracket.)

Now, largely following Pons \cite{Pons05}, we identify the condition which an infinitesimal
generator $\Igen$ must satisfy if it is to generate a gauge transformation.  Suppose that
$g(p,q)$ is an arbitrary function on phase space so that its evaluation on a phase space trajectory
 $(p(t),q(t))$ is
 $g(p(t),q(t))$. For any infinitesimal change of trajectory
  $p_i(t) \mapsto p_i(t) + \delta  p_i(t)$,
  $q^i(t) \mapsto q^i(t) + \delta   q^i(t)$ we have
  \begin{equation}\label{eqDELTAS}
    \delta \Dbd{t} g(p(t),q(t)) = \Dbd{t} \delta g(p(t),q(t)) \, .
  \end{equation}

Suppose now that  $(p_i(t),q^i(t),\La^a(t))$ and
 $(p_i(t) + \delta  p_i(t), q^i(t) + \delta   q^i(t),\La^a(t)+\delta \La^a(t))$
 are both solutions of the equations of motion, and that
 \begin{equation}
    \delta \Gt =\Pb{G}{g}(t,p(t),q(t)) \, ,
 \end{equation}
 so that $G(t,p(t),q(t))$ is the generator of the infinitesimal transformations.
   Then
   \begin{equation}
        \Dbd{t} g(p(t),q(t)) = \Pb{g}{\Hame}(p(t),q(t),\La(t))
   \end{equation}
giving
         \begin{equation}
        \delta \Dbd{t} g(p(t),q(t))= \Pbbl{G}{g}{\Hame}(t,p(t),q(t))
        + \Pb{g}{\delta   \Hame}(p(t),q(t)) \, ,
        \end{equation}
        where $\delta \Hame (p(t),q(t))= \delta \La^a(t) \,\Prim_a(p(t),q(t))\,$.
Also
 \begin{equation}
 \Dbd{t} \delta g(t,p(t),q(t))  = \Pbbr{G}{g}{\Hame} (t,p(t),q(t))
 + \Pb{\Pbdt{G}{t}}{g}(t,p(t),q(t))
 \end{equation}
so that (using (\ref{eqDELTAS}))
\begin{equation}
    \Pb{\Pbdt{G}{t}+ \Pb{G}{\Hame}+\delta\La^a \Prim_a}{g} = 0 \, .
\end{equation}
Since $g$ is an arbitrary function on phase space we must have
 \begin{equation}
    \Pbdt{G}{t}+ \Pb{G}{\Hame}+\delta\La^a \Prim_a = f(t) \, .
 \end{equation}
Following Pons, we may absorb $f(t)$ into $G(t)$ so that finally we deduce that the
requirement we can impose to determine a gauge generator is that
 \begin{equation}\label{eqPONS}
    \Pbdt{G}{t}+ \Pb{G}{\Hame} +\delta\La^a \Prim_a= 0 \,.
 \end{equation}
Also, so that the initial conditions of the trajectory are unchanged we impose $G(0)=0$.
The key condition (\ref{eqPONS}) is unchanged when the primary and secondary constraints are defined using
the covariant derivatives of the infinitesimal parameters.
We now show how generators may be constructed which satisfy
(\ref{eqPONS}) with this initial condition. (This condition will also determine the change in the Lagrangian multipliers.)  For simplicity (and because these
seem to be the examples which appear in practice) we largely consider systems
which require no more than secondary constraints, but we allow the more general
situation of equation (\ref{eqFIRSTCLASSC})
which arises when the infinitesimal
symmetry transformations of the Lagrangian involve the covariant time
derivative of the infinitesimal parameters. The extension to tertiary and higher constraints does not present new difficulties. (The BRST symmetries will in fact apply to a yet more general situation.)

 For a system with primary first
class constraints
$\Prim_a(p_i,q^i), a=1, \dots, m$ and no secondary constraints, the
result is well known:  we set
 \begin{equation}
    \Igen = \epsilon^a(t) \Prim_a(p_i(t),q^i(t))
 \end{equation}
  where each
 $\epsilon^a, a=1, \dots,m$ is an
 arbitrary function of time with $\epsilon^a(0)=0$\,. In this case,
 \begin{eqnarray}
    \lefteqn{\Pbdt{G}{t}+ \Pb{G}{\Hame} + \delta \La^a(t) \Prim_a(p_i(t),q^i(t))}\End
    &=&
    \dot{\epsilon}^a(t) \Prim_a(p_i(t),q^i(t))
       + \epsilon^a \Pb{\Prim_a}{\Hame} + \delta \La^a(t) \Prim_a(p_i(t),q^i(t)) \end{eqnarray}
 so that (\ref{eqPONS}) is satisfied if $\delta \La^a = -\dot{\epsilon}^a$.

When there are both primary constraints $\Prim_a(p,q), a=1,
\dots, m$ and secondary constraints $\Seco_a(p,q), a=1, \dots, m$ satisfying
 (\ref{eqFIRSTCLASSC})
 then we set
 \begin{equation}
    \Igen = \dot{\epsilon}^a(t) \Prim_a(p(t),q(t))- {\epsilon}^a(t) \Seco_a(p(t),q(t))
 \end{equation}
with $\epsilon^a$ arbitrary function of time with $\epsilon^a(0)=0$.
In this case
 \begin{eqnarray}
    \lefteqn{\Pbdt{G}{t}+ \Pb{G}{\Hame}(t,p(t),q(t))+\delta \La^a(t)
    \Prim_a(p_i(t),q^i(t))}\End
    &=&
    \ddot{\epsilon}^a(t) \Prim_a(p(t),q(t))
    -\dot{\epsilon}^a(t) \Seco_a(p(t),t)) \End
    &&+\dot{\epsilon}^a(t) \Pb{\Prim_a}{\Hame}(p(t),q(t))
    - {\epsilon}(t) \Pb{\Seco_a}{\Hame}+\delta \La^a(t) \Prim_a(p_i(t),q^i(t))\End
 \end{eqnarray}
 so that (\ref{eqPONS}) is satisfied if $\delta \La^a= -\Ddot{\epsilon}$. (A similar construction is possible in the covariant case.)

For the toy model with Lagrangian (\ref{eqTOYLAG}), this prescription
gives
 \begin{equation}\label{eqTOYG}
    G = \dot{\epsilon} p_1 - \epsilon p_2
 \end{equation}
so that, if $\delta \La = -\Ddot{\epsilon}$,
 \begin{eqnarray}
    \lefteqn{\Pbdt{G}{t}+ \Pb{G}{\Hame}+\delta \La p_1}\End
   &=& \ddot{\epsilon}p_1 - \dot{\epsilon} p_2
   + \dot{\epsilon} p_2+\delta \La p_1\End
   &=& 0 \,.
   \end{eqnarray}

In the following section we show that by considering the natural symplectic
structure
 on the space of paths in phase space we may recognise the gauge generators $G(t)$
which are built from the primary and secondary constraints
as moment maps. 
 \section{The symplectic geometry of paths in phase
 space}\label{secPaths}
In this section we briefly describe the symplectic structure which naturally
exists on the space of paths on a symplectic manifold (following Wurzbacher \cite{Wurzba95}) and define a
Hamiltonian group action on such a manifold together with the
associated moment map. This allows us to recognise the gauge generators constructed in the previous section in terms of the moment map of an appropriate group action.  

Let $\Man$ be a $2n$-dimensional symplectic manifold with symplectic
form $\omega$, and local canonical coordinates $p_i,x^i, i=1,\dots,n$
so that
 \begin{equation}
    \omega = \Df p_{i} \wedge \Df x^i \, .
 \end{equation}
 Also let $I=[0,T]$ be a non-empty interval in $\Real$
 and let $\Manp$ denote the set of smooth paths
 $z:I \to \Man$.   This space can be given the structure of a
 Fr\'echet manifold in a standard way, so that the tangent space
 $T_z\Manp$ at a point $z$ in $\Man$ can be identified with the space
 $\Cinf (I,z^*T\Man)$.  Thus a tangent vector $X$ at $z$ is identified with
 a vector field along $z$ in $\Man$, so that in local coordinates $X$
 may be expressed as
 \begin{equation}
    X = X_x^i(z(t)) \At{\Pbd{x^i}}{z(t)} +
    X^p_i(z(t)) \At{\Pbd{p_i}}{z(t)} \, .
 \end{equation}
It is then clear that there is a symplectic form $\Omp$ on $\Manp$ with
\begin{equation}
    \Omp(X,Y) (z(\cdot)) = X_x^i (z(\cdot)) Y^p_i (z(\cdot)) -
                           Y_x^i (z(\cdot)) X^p_i (z(\cdot)) \, ,
 \end{equation}
 so that the Poisson bracket of two functions $F$ and $G$ on $\Manp$
 takes the form
 \begin{equation}
    \Pb{F}{G} = \Fbdt{F}{x^i(z(\cdot))}\Fbdt{G}{p_i(z(\cdot))}
                    - \Fbdt{F}{p_i(z(\cdot))} \Fbdt{G}{x^i(z(\cdot))} \, .
 \end{equation}
 A detailed account of symplectic structures on loop spaces may be found in  \cite{Mokhov1998}.

 It is now possible to define a symplectic action of a Lie group on
 $\Manp$ in the usual way, and also to define such an action to be
 Hamiltonian if the infinitesimal action is obtained by Poisson bracket
 with the transpose of the moment map and there is a well-defined
 symplectic quotient.  Two simple examples will now be given, the second of which will show that
 the symplectic quotient of a Hamiltonian group action does not itself
 necessarily have the form of the space of paths in some symplectic
 manifold.  It is precisely this more general situation which is
 required to handle second class constraints.

 The first example we give of a Hamiltonian group action on $\Manp$
 arises when there is a Hamiltonian action of a Lie group $G$ on the
 symplectic manifold $\Man$ with momentum map $J$.  If $\Groupp$ denotes the group of paths
 $g:I\to G$, with group composition defined pointwise, then the action of
 $\Groupp$ on $\Manp$ defined by
 \begin{equation}
    g:z \mapsto zg \Mboxq{with} zg(t) = z(t) g(t)
 \end{equation}
is readily seen to be Hamiltonian with moment map
 \begin{equation}
    \Jp: \Manp \to \Liegdp \Mboxq{with}
    \Jp(z(\cdot))(t) = J(z(t)) \, .
 \end{equation}
 Here the dual $\Liegdp$ of the space $\Liegp$ of paths in the Lie
 algebra $\Lieg$ of $G$ is defined to be the set of smooth paths
 $\Al: I \to \Liegd$, the action of such a path on a smooth path $\eta:I \to \Lieg$
 being defined by
 \begin{equation}
    \Ipa{\Al}{\eta} = \Intt \Df t \Ipa{\Al(t)}{\eta(t)} \, .
 \end{equation}
 In this case if $B$ is the symplectic quotient of $\Man$ under the
 Hamiltonian group action of $G$ then the symplectic quotient of
 $\Manp$ is simply the space $\tilde{B}$ of paths in $B$.

 A second example of a Hamiltonian group action emerges from the toy
 example (\ref{eqTOYLAG}).  Here $\Man = \Real^4$ with canonical coordinates
 $(p_1,p_2,x^1,x^2)$ and $\Groupp=\Realp$, the group of paths
 in the additive group $\Real$.  Denoting a typical element
 $\epsilon:I \to \Real$ of $\Realp$ by $\Darg{\epsilon}$,
 the action of $\Realp$ on $\Manp$ is
 defined by
  \begin{eqnarray}
    \Darg{\epsilon}&:&(\Darg{p_1},\Darg{p_2},\Darg{x^1},\Darg{x^2}) \End
    && \mapsto (\Darg{p_1},\Darg{p_2},\Darg{x^1}+\Darg{\dot{\epsilon}},\Darg{x^2}-\Darg{\epsilon}
   ) \,.
  \end{eqnarray}
The Lie algebra $\Liegdp$ may also be identified with the space of paths in $\Real$, and the transpose $T$ of the corresponding moment map is defined by
 \begin{equation}
    T(\Ep)(t)= \dot{\Ep}(t) p_1(t)-\Ep(t) p_2(t) \, .
 \end{equation}
In this case the symplectic quotient does not take the form of the space of paths in a finite-dimensional symplectic manifold.

\section{BRST cohomology and secondary constraints}\label{secQUANTBRST}
In order to develop a BRST formalism for the more general symmetries which lead to secondary and higher constraints an alternative,
equivalent approach to phase space and the corresponding BRST operator is required; this approach will first be described for a system with no
symmetry and then extended to the general case.  This approach automatically generates, and justifies, the ghost terms which must be added to the Hamiltonian when a system has secondary constraints.

Suppose that $H$ is the Hamiltonian of a dynamical system on the symplectic manifold $\Man$ which has no symmetries (and hence no constraints).
Then consider the symplectic manifold
$\Mex=\Man\times\Real^2$ with symplectic form
  \begin{equation}
    \bar{\omega} = \omega + \Df E \wedge \Df t
 \end{equation}
 where $E$ and $t$ are global coordinates on $\Real^2$ Woodhouse \cite{Woodho}.  Now if we impose on $\Mex$ the constraint $\phi_E\equiv H-E = 0$ we are in fact using the
 constraint which corresponds to the moment map of the following action of the additive group $\Real$ on $\Mex$:

 \begin{equation}
    \Real \times \Mex \to \Mex \,,\qquad (t,(s,z)) \mapsto (s+t,\Hamflow{t}{z})
 \end{equation}
 where $\Hamflow{t}{z}$ is the flow of the Hamiltonian vector field $X_H = \Pbdt{H}{p_i}\Pbd{q^i} - \Pbdt{H}{q^i}\Pbd{p_i}$ starting from $z$, so that $\Hamflow{t}{z}$ is the solution at time $t$ of the equation of motion with initial conditions $z$.

 Now, following the standard Marsden Weinstein reduction process, we see that if $C$ is the submanifold of $\Mex$ where $H-E=0$, then $C/\Real$ is the space of trajectories $z:\Real\to \Man$ which satisfy the equations of motion for the Hamiltonian $H$.  This can of course be identified with $\Man$, given that there is a unique solution of the equations of motion starting at each point of $\Man$.

We will now construct the BRST cohomology for this constrained system and show that it does lead to the quantum dynamics of the system.  The standard construction for this single constraint gives as BRST operator
  \begin{equation}
    \Omega = \sigma(H-E)
  \end{equation}
  where $\sigma$ is the unique ghost corresponding to the basis $\Set{\Dbd{t}}$ of the Lie algebra of the additive group $\Real$. If we now choose as gauge-fixing fermion $\chi=\tau t$ where $\tau$ is the canonical dual to $\sigma$ (so that the symplectic form on the super phase space is $\bar{\omega}+\Df \tau\wedge \Df \sigma$) the gauge-fixing term is
  \begin{equation}
    \Comm{\Omega}{\chi}=(H-E)t+\tau\sigma\,.
  \end{equation}
      On quantization the term $-Et$ transforms from the Heisenberg picture (in which observables carry the dependence on time ) to the Schrodinger picture with states $\psi(q,t,\tau)$, $p_i= -i\Pbd{q^i}, \sigma=\Pbd{\tau}$. Path integration in the standard way will evaluate the supertrace of the  operator $\Exp{-Ht+\tau\sigma}$; direct calculation shows that the supertrace of $\Exp{\tau\sigma}$ is simply $1$ (as is also the case for $\Exp{\tau\sigma+A\sigma}$ provided that $A$ is independent of $\tau$). As a result carrying out $\sigma$ and $\tau$ integrals in the path integral will simply give a factor $1$, and the remaining path integral will simply be the standard path integral for the trace of the imaginary time evolution operator $\Exp{-Ht}$, as of course would be expected.

    At this stage we have regained a standard result by an apparently pointlessly roundabout route; it will now be shown that this approach can be extended to handle systems whose symmetries lead to secondary and higher order constraints, deriving the extended Hamiltonian including ghost terms.

    Suppose that we  consider the system whose classical symmetries were analysed in Section \ref{secPONS}.  Recall that this system has Hamiltonian $H(p,q)$, primary constraints $ \Prim_a$ and secondary constraints $\Seco_a$, with infinitesimal generators of classical symmetries taking the form $G(t)=\dot{\epsilon}^a(t)\Prim_a+ \epsilon(t)^a\Seco_a$, corresponding to a group action of $\Groupp$ on $\Manp$.  This group acts on the trajectories of the Hamiltonian vector field $X_H$, and so we can consider the corresponding BRST procedure.  Our parametrisation of the space of these trajectories as the quotient space of all paths in $\Man$ under the action of $\Real$ generated by $H-E$ means that we identify trajectories by their position at some arbitrary but fixed time $t_0$. Because at the particular time $t_0$ the parameters $\epsilon^a(t_0)$  and $\dot{\epsilon}^a(t_0)$ are all independent we effectively add the constraints $\Prim_a$ and $\Seco$ each with its own ghosts and can construct the BRST operator in the standard way.  At this stage it seems simpler to abandon a notation which distinguishes between primary and secondary constraints and simply consider constraints $\Cons_a$ and relations
        \begin{equation}
        \Pb{\Cons_a}{\Cons_b}=C_{ab}^c \Cons_c\,,
        \qquad \Pb{H}{\Cons_a} = V_a^b \Cons_b
    \end{equation}
    where each $C_{ab}^c$ is a constant but $V_a^b$ will in general depend on $p,q$.
    The corresponding quantum BRST operator is then
    \begin{equation}\label{eqBRSTfull}
        \Omega=\Omega_H+\Omega_G
    \end{equation}
    with $\Omega_H=\sigma (H(p,q)-E)+\sigma \eta_a \pi_b V_a^b$ and $\Omega_g=\eta^a\Cons_a+\Texthalf C_{ab}^c \eta^a \eta^b \psi_c$.
    (When $V_a^b$ is not constant a slightly more complicated BRST operator may be required.)

The BRST quantization is completed by choosing gauge fixing fermion $\chi=\chi_H + \chi_G$
where $\chi_H=\tau t$ and $\chi_G$ is chosen to have the necessary features of a gauge-fixing fermion for $\Omega_G$ \cite{GFBFVQ}.  Using the $Et$ term as before to move to the Schr\"odinger picture, we obtain the quantum Hamiltonian
 \begin{equation}
    \Comm{\Omega}{\chi}= H\,t +V_{a}^b \eta^a \pi_b t +\sigma\tau +A \sigma
    +\Comm{\Omega_G}{\chi_G}
 \end{equation}
 where $A$ is independent of $\tau$.  Once again we can integrate out $\sigma$ and $\tau$, so that the remaining path integral will evaluate the supertrace of the evolution operator
 $\Exp{H\,t + V_{a}^b \eta^a \pi_b t +\Comm{\Omega_G}{\chi_G}}$.  This is a standard result, but the arguments we have given fully justify the addition of terms $V_{a}^b \eta^a \pi_b$ to the classical Hamiltonian, rather than simply including them in order to obtain a Hamiltonian which commutes with $\Omega_G$.

To end this section this procedure will be carried out for the toy model with Lagrangian (\ref{eqTOYLAG}).  Using (\ref{eqTOYG}) and (\ref{eqBRSTfull}) we see that in terms of two ghosts $\eta^1,\eta^2$ the BRST charge is
   \begin{equation}
    \Omega=\Omega_G + \Omega_H
  \end{equation}
  where
  \begin{eqnarray}
    \Omega_H &=& \sigma(H-E)+\sigma \eta^1 \pi_2 \\
    \Omega_G &=& \eta^1 p_1 + \eta^2 p_2 \, .
  \end{eqnarray}
 In order to construct a gauge-fixing term for this theory we bring in non-minimal gauge fields and symmetries as first introduced by Fradkin and Vilkovisky \cite{FraVil2}. These fields can be interpreted as giving dynamics to the Lagrange multiplier, or alternatively simply introducing extra degrees of freedom which are then eliminated by a further symmetry.  When no secondary constraints are involved a new field $l^a$ with conjugate momentum $k_a$ is introduced for each constraint $\phi_a$, while the Lagrangian is unchanged, so that each $k_a$ is constrained to be zero. The corresponding BRST operator then has additional terms $\kappa^ak_a$ where
 $\kappa^a$ is the ghost corresponding to the constraint $k_a=0$. In the case of secondary constraints extra terms must be added to the Hamiltonian.  For the toy model there are two constraints and so two new fields $l^1,l^2$ with conjugate momenta $k_1,k_2$  are introduced and an extra term $\frac{1}{\epsilon}l^1 k_2$ is  added to the Hamiltonian, so that $k_1$ is a primary constraint and $k_2$ a secondary constraint.  (Here $\epsilon$ is a constant factor.)  The corresponding BRST charge is then

   \begin{equation}
    \Omega_{NM}=\Omega_H + \Omega_G +\Omega_K
    \Mboxq{where} \Omega_K=
    + \frac{1}{\epsilon}\sigma l^1k_2 +
    \frac{1}{\epsilon}\sigma\kappa^1\lambda_2 +\kappa^1k_1+\kappa^2 k_2
   \end{equation}
   and $\La_1$ and $\La_2$ are the ghost momenta conjugate to $\kappa^2$ and $\kappa^2$ resepectively.

   It will now be shown that choosing the gauge-fixing term 
   \begin{equation}
    \chi_{NM}= \tau t + \pi_1 l^2 + \frac{1}{\epsilon} q^1 \La_2
   \end{equation}
   leading to the gauge-fixed Hamiltonian
   \begin{eqnarray}
     H_{NM} &=& H_c + \frac{1}{\epsilon} l^1k_2 + \eta^1\pi_2  +\frac{1}{\epsilon} \kappa^1\La_2\End
      && +\pi_1 \kappa^2 + \frac{1}{\epsilon}\La_2\eta^1
      +\frac{1}{\epsilon}q^1 k_2 + l^2 p_1
   \end{eqnarray}
   after the $Et$ term has been used as before and integration over $\sigma$ and $\tau$ has been carried out. Inspecting this Hamiltonian we see that it has the standard form to ensure that the supertrace of $\Exp{i H_{NM} t}$ will project to the trace over physical states \cite{GFBFVQ}.  This may be seen from the fact that 
   $H_{NM}=H_D + [\pi_1 l^2 + \frac{1}{\epsilon} q^1 \La_2,\Omega_G+\Omega_K]$ where 
   $H_D=H_c + \frac{1}{\epsilon} l^1k_2 + \eta^1\pi_2  +\frac{1}{\epsilon} \kappa^1\La_2$ and by construction $[H_D,\Omega_G + \Omega_K]=0$ together with the fact that $[\pi_1 l^2 + \frac{1}{\epsilon} q^1 \La_2,\Omega_G+\Omega_K]=\pi_1 \kappa^2 + \frac{1}{\epsilon}\La_2\eta^1
      +\frac{1}{\epsilon}q^1 k_2 + l^2 p_1$ which is the standard form for the quartet mechanism \cite{HenTei}.

  For completeness, it will now be shown that the corresponding path integral reduces to the form given in \cite{DreFisGreHen} after integration over some of the variables. The path integral takes the form
   \begin{eqnarray}
     Z &=& \int  \Dc p_1 \Dc q^1 \Dc p_2 \Dc q^2 
     \Dc k_1 \Dc l^1 \Dc k_2 \Dc l^2 
     \Dc \pi_1 \Dc \eta^1 \Dc \pi_2 \Dc \eta^2 \Dc \kappa^1 \Dc\La_1 \Dc\kappa^2\Dc\La_2\End
      &&\exp\biggl(\int_0^t i(p_1 \dot{q}^1+p _2\dot{q}^2+k_1\dot{l}^1+k_2\dot{l}^2+
      \pi_1\dot{\eta}^1 + \pi_2\dot{\eta} ^2+\kappa^1\dot{\La}_1+\kappa^2\dot{\La}_2)\End
      &&\qquad\qquad\qquad\qquad+H_{NM})\Df t\biggr)\,.
   \end{eqnarray}
   To evaluate this path integral we use a technique first introduced by Fradkin and Vilkovisky \cite{FraVil2} and scale $k_2 \to \Eps k_2$ 
   and $\La_2 \to \Eps \La_2$ (which leaves the measure unchanged) and let $\epsilon \to 0$. The integrand then becomes
   \begin{eqnarray}
  && \exp\biggl(\int_0^t i(p_1 \dot{q}^1+p _2\dot{q}^2+k_1\dot{l}^1+
      \pi_1\dot{\eta}^1 + \pi_2\dot{\eta} ^2+\kappa^1\dot{\La}_1)\End
      &&+\frac12 p_2^2-q^1p_2 + l^1k_2+\eta^1\pi_2 +\kappa^1\La_2
      +\pi_1\kappa^2+\La_2\eta^1 +q^1k_2+l^2p_1)\Df t\biggr)\,.\end{eqnarray}
Integrating out $l^1$ gives a factor $\delta(k_2-\dot{k}_1)$, while integrating out $l^2$ gives a factor $\delta(p_1)$.  On the ghost side, integrating out $\pi_2$, $\kappa^1$ and $\kappa^2$ give factors $\delta(\dot{\eta}^2+\eta^1), \delta(\La_2-\dot{\La}_1)$ and $\delta(\pi^1)$ respectively.  Integrating out 
$p_1,p_2,k_2,\pi_1,\eta^1$ and $\La_2$ is then straightforward, reducing the path integral to the form
  \begin{eqnarray}
     Z &=& \int \Dc q^1  \Dc q^2   \Dc k_1  \Dc \eta^2  \Dc\La_1 \End
      &&\exp\biggl(\int_0^t i\Lrb{ \frac12 (q^1-\dot{q}^2)^2+k_1\dot{q}^1 +\dot{\La}_1\dot{\eta}^2}\Df t\biggr)\,.
   \end{eqnarray}
  in agreement with (4.16) of \cite{DreFisGreHen} if $k_1$ is replaced by $b$, $\La_1$ by $\bar{C}$ and $\eta^2$ by $C$.
\section{Examples}\label{secEXAMPLES}
In this section we present some examples to illustrate how  primary and secondary constraints are combined in a single generators of gauge transformations that satisfies Pons' condition \eqref{eqPONS}. These examples are the relativistic free particle, pure Yang-Mills theory and the bosonic string, all of which have secondary constraints. The first of these is an abelian theory, whereas the second and third are non-abelian field theories.
For each example, the action is presented. Then, following Dirac's algorithm, the constraints are deduced and a generator of the gauge transformations is constructed which is shown to satisfy the required condition.

\subsection{Relativistic free particle}
This system corresponds to a (0+1)-dimensional gravity theory with scalar fields. The degrees of freedom are  particle coordinates $x^{\mu}$ and an einbein $e$, each of which is a function of a single proper time variable $\tau$. The action for this system is given by
\begin{equation}\label{relparticle}
S=\int \ud\tau \; \Lagr =\frac{1}{2}\int \ud\tau \left(\dfrac{\dot{x}_{\mu}(\tau)\dot{x}^{\mu}(\tau)}{e(\tau)}-\um^{2}e(\tau) \right)\; ,
\end{equation}
where a dot over a variable indicates a derivative with respect to proper time $\tau$.
From the action \eqref{relparticle}, the conjugate momenta to $e$ and $x^{\mu}$ are respectively
\begin{equation}
P_{e}=\dfrac{\partial \Lagr}{\partial\dot{e}}=0 \Mboxq{and}
P_{\mu}=\dfrac{\partial \Lagr}{\partial\xu }=\dfrac{\xd}{e}\;,
\end{equation}
so that we have the primary constraint
\begin{equation}
\Prim=P_{e}=0\;\;\; .
\end{equation}
The extended hamiltonian is then
\begin{equation}
\Hame=H+v^{1}\phi_{1}\;\;\; ,
\end{equation}
where $v^{1}$ is an arbitrary function of $\tau$, and $H$ is the canonical Hamiltonian which takes the form
\begin{equation}
H=\frac{1}{2}e(P_{\mu}P^{\mu}+\um^{2})\; .
\end{equation}
Since $\Pb{H}{\Prim}=\frac{1}{2}\left( P_{\mu}P^{\mu}+\um^{2}\right)$ we
have the secondary constraint
\begin{equation}
\Seco=\frac{1}{2}\left( P_{\mu}P^{\mu}+\um^{2}\right) =0\;\;\; .
\end{equation}
This new constraint satisfies $\Pb{H}{\Seco}=0$ and so there are no further constraints.

Finally we construct the gauge generator
 $G(t)=\dot{\Ep}\Prim+ \Ep\Seco$ which satisfies  \eqref{eqPONS} 
provided that $\delta\La =\Ddot{\Ep}$.


\subsection{Yang-Mills}
This is a very well known gauge theory where the fundamental fields are gauge potentials $A^{a}_{\mu}, a=1, \dots, \Mbox{\ dim\ } G$ where $G$ is the gauge group of the theory. The Lagrangian density for this system is
\begin{equation}
\mathcal{L}=\frac{1}{4}F_{\mu\nu}^{a}F^{\mu\nu}_{a}
\end{equation}
where $F_{\mu\nu}^{a}=\partial_{\mu}A_{\nu}^{a}-\partial_{\nu}A_{\mu}^{a}
-gf^{a}_{\;bc}A_{\mu}^{b}A_{\nu}^{c}$ with $g$ the coupling constant and $f^{a}_{\;bc}$ the structure constants of the Lie algebra of the gauge group $G$.
The conjugate momenta are
\begin{eqnarray}
&&\pi_{a}^{0}=\frac{\partial \mathcal{L}}{\partial\dot{A}_{0}^{a}}=0\\
&&\pi_{a}^{i}=\frac{\partial \mathcal{L}}{\partial\dot{A}_{i}^{a} }=-F_{i0}^{a}
\end{eqnarray}
so that there are primary constraints
\begin{equation}
\phi^{\left( 1\right) }_{a}=\pi_{a}^{0}
\end{equation}
while the canonical Hamiltonian  is
\begin{equation}
 H=\int \Df^3 \mathbf{x} \Lrb{\frac{1}{2}\pi_{a}^{i}(\mathbf{x},t)\pi_{a}^{i}
 (\mathbf{x},t)-A_{0}^{a}(\mathbf{x},t)
 \mathcal{D}_{i}\pi_{a}^{i}(\mathbf{x},t)
 +\frac{1}{4}F_{ik}^{a}(\mathbf{x},t)F^{ik}_{a}(\mathbf{x},t) }
\end{equation}
where the covariant derivative is defined by
$\Dc_{\mu}\pi^{\nu}_a
 =\partial_{\mu} \pi^{\nu}_a + g A_{\mu}^b f_{ab}^c\pi^{\nu}_c\,.$ and spacetime has been taken to have four dimensions.

Since $\Pb{H}{\Prim}=\mathcal{D}_{i}\pi_{a}^{i}$
we have secondary constraints
\begin{equation}
\Seco_{a}=-\mathcal{D}_{i}\pi_{a}^{i}=0
\end{equation}
These new constraints satisfy
\begin{equation}
\Pb{H}{\Seco_a}
=-gf_{abc}A_{0}^{b}\phi^{2}_{c}\approx 0
\end{equation}
and so there are no tertiary or higher constraints.  The Poisson brackets of the constraints with each other are
 \begin{eqnarray}
    \Pb{\Prim_a(\mathbf{x},t)}{\Prim_b(\mathbf{y},t)}&=&0\, , \quad\Pb{\Prim_a(\mathbf{x},t)}{\Seco_b(\mathbf{y},t)}=0 \End
    \Mboxq{and}
    \Pb{\Seco_a(\mathbf{x},t)}{\Seco_b(\mathbf{y},t)}
    &=& f_{ab}^c \Seco_c(\mathbf{y},t) \delta(\mathbf{x}-\mathbf{y}) \, .
 \end{eqnarray}

The form of the constraints suggests the generator
\begin{equation}
G(\Ep)=\left( \mathcal{D}_{0}\epsilon^{a}\right) \phi^{1}_{a}+\epsilon^{a}\phi^{2}_{a}
\end{equation}
with each $\epsilon^{a}$ an arbitrary function of time.
Now
\begin{eqnarray}
&&\partial_{0}G=\left( \partial^{2}_{0}\epsilon^{a}-gf_{abc}A^{c}_{0}\partial_{0}\epsilon^{b}\right)\phi^{1}_{a}+\left(\partial_{0}\epsilon^{a} \right)\phi^{2}_{a}  \\
\Mbox{and} &&\lbrace G,\mathcal{H}_{c}\rbrace =-\partial_{0}\epsilon^{a}\phi^{2}_{a}\\
\Mbox{so that}&&\partial_{0}G+\lbrace G,\mathcal{H}_{c} \rbrace =\left(  \mathcal{D}_{0}\partial_{0}\epsilon^{a}\right) \phi^{1}_{a}
\end{eqnarray}
and thus (\ref{eqPONS}) is satisfied if $\delta\La^a=\mathcal{D}_{0}\partial_{0}\epsilon^{a}$\,. Direct calculation shows that
\begin{equation}
    \Pb{G(\Ep_1)}{G(\Ep_2)}=G(\Ep_3)
\end{equation}
where $\Ep_3^a = \Ep_1^b \Ep_2^c f^a_{bc}$ so that the symmetry group is  the space of maps of spacetime into $G$, but the group action is not pointwise.

\subsection{Bosonic String}
Relativistic free strings propagating in an arbitrary D-dimensional fixed background
space-time can be described with the Polyakov action \cite{Polyak1}
 \begin{equation}
    S[\Ga^{ab},X^{\mu}] =
    \alpha \int d^2 \xi \sqrt{-\Ga} \Ga^{ab} \partial_a X^\mu \partial_b X^\nu
    g_{\mu\nu}(X)
 \end{equation}
 where $X^{\mu}$ are back-ground spacetime coordinates, $g_{\mu\nu}$ is the background spacetime metric,  $\Ga^{ab}$ is the world sheet metric and $\xi$ stands for the two worldsheet coordinates $\tau$ and $\sigma$. When $\Ga^{ab}$ is taken in the ADM form \cite{ArnDesMis}
 \begin{equation}
    \Ga^{ab}= \Matt{-\frac{1}{N^2}}{\frac{\La}{N^2}}{\frac{\La}{N^2}}
    {\frac{1}{\chi} -\frac{\La^2}{N^2}}
 \end{equation}
 the Lagrangian density in the Polyakov action takes the form \cite{MonVer}
 \begin{equation}
\mathcal{L}=\frac{\alpha\epsilon\sqrt{\chi}}{N}\left(-\Xd{\mu}\Xd{\nu}\gdmn+2\lambda\Xd{\mu}\Xp{\nu}\gdmn+ \left(\frac{N^{2}}{\chi}-\lambda^{2}\right)\Xp{\mu}\Xp{\nu}\gdmn\right)
\end{equation}
where
\begin{equation}
\Xd{\mu}=\fpartial{X^{\mu}}{\tau};\:\:\:\Xp{\mu}=\fpartial{X^{\mu}}{\sigma} \, .
\end{equation}

The conjugate momenta to $X^{\mu}$ are:
\begin{equation}
P_{\mu}=-\frac{2\alpha\epsilon\sqrt{\chi}}{N}\Xd{\nu}\gdmn+\frac{2\alpha\epsilon\lambda\sqrt{\chi}}{N}\Xp{\nu}\gdmn
\end{equation}
while those of $\lambda ,N,\chi $ are
\begin{equation}
P_{\lambda}=P_{N}=P_{\chi}=0
\end{equation}
giving the primary constraints
\begin{equation}
\prim{\lambda}=P_{\lambda}\; ,\;\;\;  \prim{N}=P_{N}\; \Mbox{and}\;\;\;  \prim{\chi}=P_{\chi} \, .
\end{equation}

The canonical Hamiltonian  is
\begin{equation}
H_{c}(\tau)=
\int \Df\sigma \quad
\Hamden(\chi(\sigma,\tau),N(\sigma,\tau),
\lambda(\sigma,\tau),X^{\mu}(\sigma,\tau),P_{\mu}(\sigma,\tau))
\end{equation}
where
\begin{equation}
\Hamden =
\Lrb{-\frac{N}{4\alpha\epsilon\sqrt{\chi}}\left(P_{\mu}P_{\nu}\gumn
+ 4\epsilon^{2}\alpha^{2}\Xp{\mu}\Xp{\nu}\gdmn\right)+\lambda\Xp{\mu}P_{\mu}} \,.
\end{equation}
Taking Poisson brackets with the primary constraints gives
\begin{eqnarray}
&&\Pb{H_c}{\Prim_{\La}}=-\Xp{\mu}P_{\mu} \\
&&\Pb{H_c}{\prim{N}}=-\frac{1}{4\alpha\epsilon\sqrt{\chi}}\left(P_{\mu}P_{\nu}\gumn+4\alpha^{2}\epsilon^{2}\Xp{\mu}\Xp{\nu}\gdmn \right)\\
&&\Pb{H_c}{\prim{\chi}}=-\frac{N}{8\alpha\epsilon\chi^{3/2}} \left(P_{\mu}P_{\nu}\gumn+4\alpha^{2}\epsilon^{2}\Xp{\mu}\Xp{\nu}\gdmn \right)
\end{eqnarray}
 and hence we have secondary constraints
\begin{eqnarray}
&&\secu{D}=\Xp{\mu}P_{\mu}\\
&&\secu{H}=P_{\mu}P_{\nu}\gumn+4\alpha^{2}\epsilon^{2}\Xp{\mu}\Xp{\nu}\gdmn \, .
\end{eqnarray}

If we define
\begin{equation}
    \Seco_{H}(\Ep)= \int \Df\sigma \Ep(\sigma) \Seco_{H}(\sigma)
    \Mboxq{and} \Seco_{D}(\Ep)= \int \Df\sigma \Ep(\sigma) \Seco_{D}(\sigma)
\end{equation}
direct calculation shows that \cite{MonVer}
\begin{eqnarray}
\lbrace\Seco_{D}(\Ep),\Seco_{D}(\Al)\rbrace&=&\secu{D}(\Ka)\\
\lbrace\Seco_{D}(\Ep),\Seco_{H}(\Al)\rbrace&=&\secu{H}(\Ka)\\
\lbrace\Seco_{H}(\Ep),\Seco_{H}(\Al)\rbrace&=&\secu{D}(\Ka)
\end{eqnarray}
where $\Ka= \Ep'\Al-\Al'\Ep$.

Since
\begin{equation}
H_c=-\rho \Seco_{H}+\La\Seco_{D}
\end{equation}
(where $\rho=\frac{N}{4\alpha\epsilon\sqrt{\chi}}$) there are no tertiary or higher constraints, which suggests a generator of the form
\begin{eqnarray}
G(\tau)&=&\int \Df\sigma \Bigl( \epsilon^{\lambda}(\sigma,\tau)\prim{\lambda}(\sigma,\tau)
+\epsilon^{N}(\sigma,\tau)\prim{N}(\sigma,\tau)
+\epsilon^{\chi}(\sigma,\tau)\prim{\chi}(\sigma,\tau)\End
&&\qquad+\epsilon^{D}(\sigma,\tau)\secu{D}(\sigma,\tau)
+\epsilon^{H}\secu{H}(\sigma,\tau)\Bigr)
\end{eqnarray}
with each $\epsilon$ an arbitrary function of $\tau$ and $\sigma$. By substituting this generator in the equation \eqref{eqPONS}, we find
\begin{eqnarray}
\lefteqn{\fpartial{G}{\tau}+\lbrace G, H\rbrace}\End
 &=& \int\Df\sigma \Biggl[\dot{\epsilon}^{\lambda}\prim{\lambda}+\dot{\epsilon}^{N}\prim{N}+\dot{\epsilon}^{\chi}\prim{\chi} \End
&+&\left(\dot{\epsilon}^{D}-\epsilon^{\lambda}+
 (\lambda'\epsilon^{D}-\lambda(\epsilon^{D})')
-(\rho'\epsilon^{H}-\rho(\epsilon^{H})') \right)\secu{D} \End
&+& \left(\dot{\epsilon}^{H}
+\frac{\rho}{N}\epsilon^{N}
-\frac{\rho}{2\chi}\epsilon^{\chi}
-(\rho'\epsilon^{D}-\rho(\epsilon^{D})')
+(\lambda'\epsilon^{H}+\lambda(\epsilon^{H})') \right) \secu{H}\Biggr]\,.\End
\end{eqnarray}
so that the parameters must obey the relationships
\begin{eqnarray}
\epsilon^{\lambda}&=&\dot{\epsilon}^{D}
+(\lambda'\epsilon^{D}-\lambda(\epsilon^{D})')
+(\rho'\epsilon^{H}-\rho(\epsilon^{H})')\\
\epsilon^{N}&=& \frac{N}{2\chi}\epsilon^{\chi}
-\frac{N}{\rho}\left( \dot{\epsilon}^{H} -(\lambda'\epsilon^{H}+\lambda(\epsilon^{H})')
+(\rho'\epsilon^{D}-\rho(\epsilon^{D})')\right) 
\end{eqnarray}
and the generator is
\begin{eqnarray*}
G&=&\int\Df\sigma\Biggl[\left( \dot{\epsilon}^{D}
+(\lambda'\epsilon^{D}-\lambda(\epsilon^{D})')
+(\rho'\epsilon^{H}-\rho(\epsilon^{H})')\right) \prim{\lambda}\End
&&+\left( \frac{N}{2\chi}\epsilon^{\chi}
-\frac{N}{\rho}\left( \dot{\epsilon}^{H} -(\lambda'\epsilon^{H}+\lambda(\epsilon^{H})')
+(\rho'\epsilon^{D}-\rho(\epsilon^{D})')\right)\right)\prim{N}\\
&&+\epsilon^{\chi} \prim{\chi}
+\epsilon^{D}\secu{D}+\epsilon^{H}\secu{H}\Biggr]\,.
\end{eqnarray*}
%

 \end{document}